\documentclass[a4paper,aps,pre,10pt,
 twocolumn,
 notitlepage,
 showpacs,floatfix,nofootinbib,
 superscriptaddress
 ]{revtex4-1}

 \usepackage{subfigure}
 \usepackage{amssymb}
 \usepackage{amsfonts}
 \usepackage{amsmath}
 \usepackage{amsthm}
 \usepackage{epsfig}
 \usepackage{float}
 \usepackage[usenames,dvipsnames]{color}
 \usepackage[latin1]{inputenc}

 \usepackage{graphics,graphicx,color,subfigure}
\usepackage{longtable}

 \usepackage[hidelinks,unicode=true]{hyperref}
 \hypersetup{colorlinks=true,
 	linkcolor=blue,
 	urlcolor=blue,
 	citecolor=blue,
 	pdfhighlight=/N
 }

 \usepackage{comment}
 \usepackage[normalem]{ulem}

\newcommand{\av}[1]{\langle {#1} \rangle}
\newcommand{\ave}[1]{\langle {#1} \rangle_\text{e}}

\graphicspath{./}

\newcommand{\kmax}{k_\text{max}}
\newcommand{\kmin}{k_\text{min}}

\begin{document}

\title{Spectral properties and the accuracy of mean-field approaches
for epidemics on correlated networks }

\author{Diogo H. Silva}
\affiliation{Departamento de F\'{\i}sica, Universidade Federal de
  Vi\c{c}osa, 36570-900 Vi\c{c}osa, Minas Gerais, Brazil}
\author{Silvio C. Ferreira}
\affiliation{Departamento de F\'{\i}sica, Universidade Federal de
  Vi\c{c}osa, 36570-900 Vi\c{c}osa, Minas Gerais, Brazil}
\affiliation{National Institute of Science and Technology for Complex
  Systems, Brazil}
\author{Wesley  Cota}
\affiliation{Departamento de F\'{\i}sica, Universidade Federal de
	Vi\c{c}osa, 36570-900 Vi\c{c}osa, Minas Gerais, Brazil}

\author{Romualdo Pastor-Satorras} \affiliation{Departament de
  F\'{\i}sica, Universitat Polit\`{e}cnica de
  Catalunya, Campus Nord B4, 08034 Barcelona, Spain}
\author{Claudio Castellano} \affiliation{Istituto dei Sistemi
  Complessi (ISC-CNR), Via dei Taurini 19, I-00185 Roma,Italy}

\begin{abstract}
  We present a comparison between stochastic simulations and mean-field theories
  for the epidemic threshold of the susceptible-infected-susceptible (SIS) model
  on correlated networks (both assortative and disassortative) with power-law
  degree distribution $P(k)\sim k^{-\gamma}$. We confirm the vanishing of the
  threshold regardless of the correlation pattern and the degree exponent
  $\gamma$. Thresholds determined numerically are compared with quenched
  mean-field (QMF) and pair quenched mean-field (PQMF) theories.  Correlations
  do not change the overall picture: QMF and PQMF provide estimates that are
  asymptotically correct for large size for $\gamma<5/2$, while they only
  capture the vanishing of the threshold for $\gamma>5/2$, failing to reproduce
  quantitatively how this occurs.  For a given size, PQMF is more accurate.  We
  relate the variations in the accuracy of QMF and PQMF predictions with changes
  in the spectral properties (spectral gap and localization) of standard and
  modified adjacency matrices, which rule the epidemic prevalence near the
  transition point, depending on the theoretical framework. We also show that,
  for $\gamma<5/2$, while QMF provides an estimate of the epidemic threshold
  that is asymptotically exact, it fails to reproduce the singularity of the
  prevalence around the transition.
\end{abstract}

\maketitle

\section{Introduction}
\label{sec:intro}

Metabolic chains of protein interactions~\cite{Vazquez2003}, collaborations
among scientists, co-starring in a movie~\cite{Newman2014}, or person-to-person
contacts~\cite{Cattuto2010}, are all examples of interacting systems that can be
modeled using complex networks~\cite{Newman2014}. A large number of networks
representing real systems show a heavy-tailed degree distribution described by a
power-law, $P(k)\sim k^{-\gamma}$ \cite{Barabasi:1999,barabasi02}, usually with
strong levels of correlations~\cite{Pastor-Satorras2001,Newman2002}. Degree
correlations are encoded in the conditional probability $P(k'|k)$ that a vertex
of degree $k$ is connected to a vertex of degree
$k'$~\cite{Pastor-Satorras2001}. Technological networks, such as the Internet,
show in general disassortative mixing~\cite{Pastor-Satorras2001,Newman2002},
i.e., vertices of large degree tend to be connected with those of small degree,
and vice-versa.  Assortative mixing occurs in social networks, where connections
preferentially occur among vertices exhibiting similar degree.  Since 
uncorrelated networks usually simplify theoretical approaches, they are
typical benchmarks for the investigation of dynamical processes on networks and
have been considered in many
studies~\cite{Vespignani2008,Pastor-Satorras2015b,Wang2017}.  However, the
ubiquitousness of correlations in real networks naturally calls for the
investigation of the effect of correlated interaction patterns. While the
effects of degree correlations have been considered for several dynamical
processes
\cite{Boguna2003,structured,sander,PVM_assortativity_EJB2010,PhysRevE.78.051105,Kenah2011},
a full understanding of their effects on the performance of theoretical
approaches is still missing.

A basic approach to investigate dynamical processes on networks is the
heterogeneous mean-field (HMF) theory, in which degree heterogeneity
and correlations are taken into account through the distributions
$P(k)$ and $P(k|k')$,
respectively~\cite{Pastor-Satorras2001b,Pastor-Satorras2015b,Vespignani2008,barabasi2016network}. A
more refined approach is provided by the quenched mean-field theory
(QMF)~\cite{YangWang,Chakrabarti2008,Castellano2010}, which considers
the full topology as described by the unweighted adjacency matrix
(defined as $A_{ij}=1$ if vertices $i$ and $j$ are connected and
$A_{ij}=0$ otherwise) and thus takes into account the detailed
connectivity structure.

A crucial question in this context is the ability of theories to accurately
predict the epidemic threshold of the susceptible-infected-susceptible (SIS)
dynamics, the most basic epidemic process with an absorbing-state
phase-transition~\cite{Castellano2010,Ferreira2012,Goltsev2012,Lee2013,Boguna2013,Mata2013,Mata2015,Castellano2017}.
For random uncorrelated networks, such as those created according to the
uncorrelated configuration model~\cite{Catanzaro2005}, when $\gamma<5/2$ the two
theories tend to agree, predicting a vanishing threshold as the network size
diverges~\cite{Ferreira2012}.  For $\gamma>3$ instead, QMF theory correctly
predicts again the asymptotic vanishing of the epidemic
threshold~\cite{Chatterjee2009}, while HMF fails, predicting the existence of a
finite threshold.  In spite of being qualitatively correct, QMF theory is
however not able to accurately predict the effective finite-size epidemic
threshold in this case~\cite{Boguna2013}.  A further quantitative improvement of
the QMF theory has been achieved in Ref.~\cite{Mata2013} [hereafter called of
pair QMF (PQMF) theory] by means of the explicit inclusion of pairwise dynamical
correlations~\cite{Gleeson2011,Gleeson2013,Cator2012}

In this work, we investigate the ability of the aforementioned
approaches (HMF, QMF, PQMF) to quantitatively predict the value of the
epidemic threshold for both uncorrelated and correlated networks
generated using the Weber-Porto model~\cite{Weber2007} and for
real-world topologies.
We find that correlations do not change qualitatively the
scenario found on uncorrelated networks.  The epidemic threshold
vanishes asymptotically with the system size for both assortative and
disassortative correlations.  For $\gamma<5/2$, both QMF and PQMF seem
to provide an asymptotically exact estimate of the numerical
threshold, while they are only qualitatively correct for $\gamma>5/2$.
As in the case of uncorrelated networks~\cite{Mata2013}, PQMF
outperforms the other theories.
The amplitude of the discrepancies between numerics and theory is correlated
with violations of the assumptions underlying them, revealing that both
theories tend to be more accurate if the principal
eigenvector of the (effective) adjacency matrix is not strongly localized
or the spectral gap is large.
The same scenario is found to hold when SIS dynamics is considered on a set of
real-world topologies.  In addition, we analyze the singularity of the
prevalence near the transition point through the critical exponent $\beta$,
defined as $\rho\sim (\lambda-\lambda_{c})^{\beta}$.  Interestingly, we find
that for $\gamma<5/2$ even if QMF provides an asymptotically exact estimate of
the position of the epidemic threshold, the QMF prediction for the prevalence
exponent, $\beta^\text{(QMF)}=1$~\cite{VanMieghem2012,Goltsev2012}, is correct
only not too close to the transition.

The rest of the paper is organized as follows.  Section~\ref{sec:simulations}
describes the models used to generate correlated heavy-tailed networks, the
implementation of the SIS model, and the theoretical approaches. The comparison
between simulations and theory on synthetic and real networks in presented  in
Sec.~\ref{sec:results}.  Our conclusions are drawn in Sec.~\ref{sec:conclusion}.
Appendix~\ref{app:real}, summarizing the properties of the investigated
real networks, complements the paper.
\section{Models and Methods}
\label{sec:simulations}

\subsection{Weber-Porto configuration model}
\label{subsec:WP}

The degree correlations encoded in the conditional probability $P(k'|k)$ can be
more easily interpreted by the simple metrics of the average degree of the
nearest-neighbors as a function of the vertex degree~\cite{Pastor-Satorras2001},
defined as
\begin{equation}
\kappa_\text{nn}(k)=\sum_{k'=\kmin}^{\kmax} k'P(k'|k),
\end{equation}
where $\kmin$ and $\kmax$ are the lower and upper cutoffs of the degree
distribution. If $\kappa_\text{nn}(k)$ increases or decreases with $k$, the
networks are assortative or disassortative, respectively. In the case of
uncorrelated networks we have~\cite{Pastor-Satorras2001}
\begin{equation}
P(k'|k)=P_\text{e} (k')=k'P(k')/\av{k}
\end{equation}
which implies that $\kappa_\text{nn}=\av{k^2}/\av{k}=\ave{k}$ does not depend on
$k$. We use here the edge distribution average
$\ave{A(k)}=\sum_kA(k)P_\text{e}(k)$ where $P_\text{e}(k)$ is the probability
that an edge ends on a vertex of degree $k$.

We are interested in heavy-tailed networks with degree distribution $P(k)\sim
k^{-\gamma}$ and correlation given by $\kappa_\text{nn}(k)\sim k^{\alpha}$.
These networks can be generated using an algorithm proposed by Weber and
Porto~\cite{Weber2007}, hereafter called Weber-Porto configuration model (WPCM).
The degree of each vertex is drawn according to the degree distribution $P(k)$
and initially each node has $k$ unconnected stubs. Two stubs are randomly chosen
and connected with probability
\begin{equation}
P_\text{link}(q',q)=\frac{f(q',q)}{f_\text{max}},
\label{eq:plink}
\end{equation}
where $q$ and $q'$ are the respective degrees of the
chosen vertices and $f_\text{max}$  is
the maximum value of
\begin{equation}
f(q,q')=1+\frac{(\kappa_\text{nn}(q)-\ave{k})(\kappa_\text{nn}(q')-\ave{k})}{
	\ave{k\kappa_\text{nn}}-\ave{k}^{2}},
\label{eq:fq_qp}
\end{equation}
computed over the whole network.  Self- and multiple connections are
forbidden. In the absence of degree correlations, we have
$\kappa_\text{nn}=\ave{k}$, implying $f(q,q')=1$ and $P_\text{link}=1$.  See
Ref.~\cite{Weber2007} for more details.
\begin{figure}[t]
\centering
\includegraphics[width=0.8\linewidth]{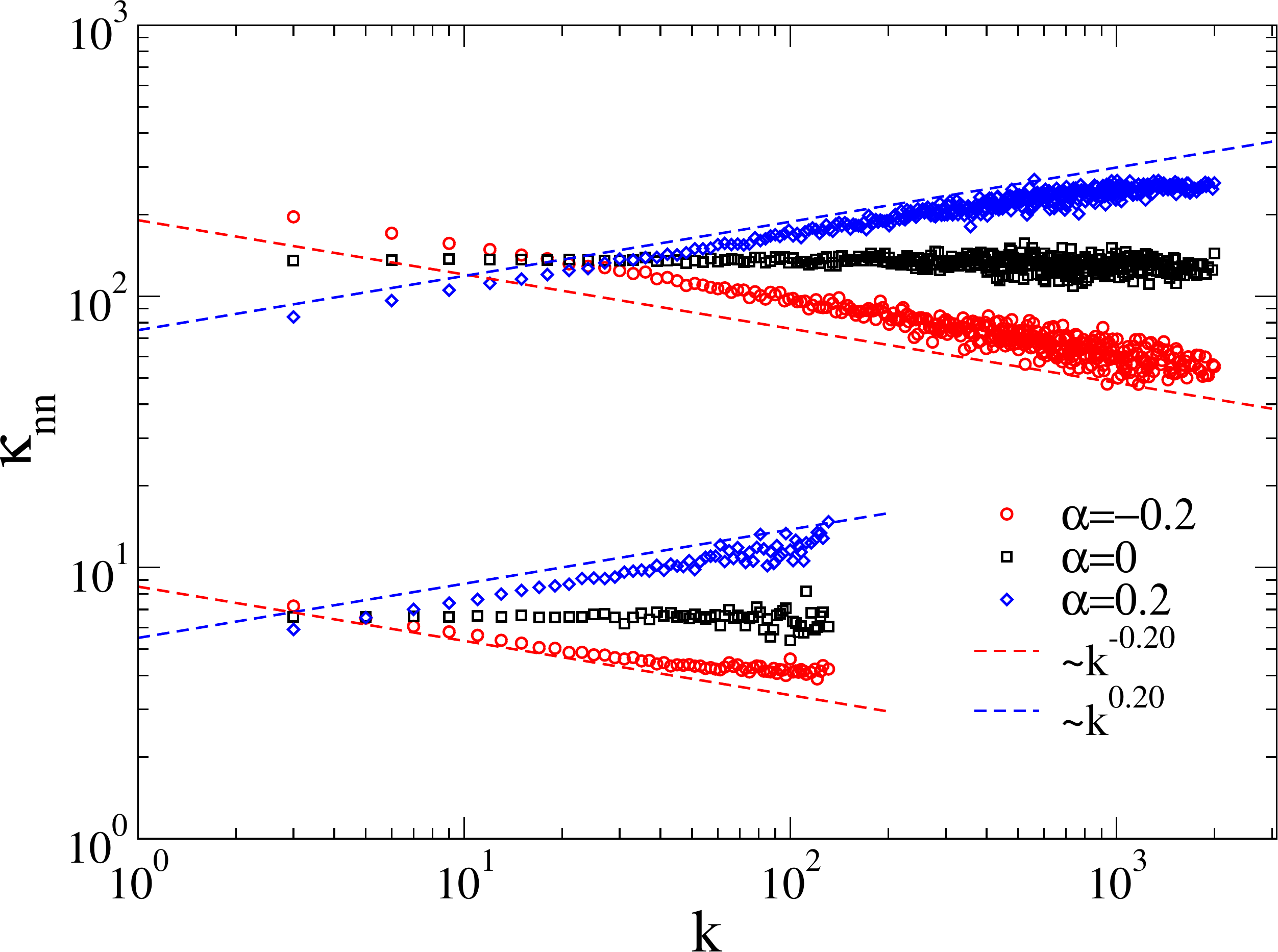}
\caption{Average degree of the nearest-neighbors as a function of the degree for
  networks built with the WPCM algorithm~\cite{Weber2007} for power-law degree
  distributions with $\gamma=2.3$ (top curves) and $\gamma=3.5$ (bottom
  curves). The network size is $N=10^{6}$, the lower cutoff is $\kmin=3$. The
  upper cutoff is given by $\kmax=2\sqrt{N}$ for $\gamma=2.3$ and $NP(\kmax)=1$
  for $\gamma=3.5$.}q
\label{fig:net}
\end{figure}

Figure~\ref{fig:net} shows $\kappa_\text{nn}$ as a function of $k$ for networks
obtained with the WPCM algorithm~\cite{Weber2007} using different values of
$\gamma$ and $\alpha$, with lowest degree $\kmin=3$.  We adopt different upper
cutoffs for the degree distribution. For $\gamma<3$, the structural cutoff
$\kmax=2\sqrt{N}$~\cite{Boguna2004} is used, while for $\gamma>3$, a rigid
cutoff is determined by the condition $NP(\kmax)=1$~\cite{Dorogovtsev2002}. The
first choice allows to enhance the effects of hubs and to approach faster the
thermodynamic limit while fulfilling the criterion $\kmax<\sqrt{\av{k} N}$
necessary to produce uncorrelated networks in the case
$\alpha=0$~\cite{Boguna2004}.  The second choice is justified by numerical
reasons explained in Sec.~\ref{sucsec:SIS}.  The predetermined scaling law
$\kappa_\text{nn}(k)\sim k^\alpha$ is very well reproduced. Small deviation for
positive or negative $\alpha$ are due to the network finite size that prevents
$\kappa_\text{nn}$ from decaying or increasing indefinitely with $k$.  The range
of the power-law behavior is extended as the network size increases.

\subsection{SIS simulations}
\label{sucsec:SIS}

In the SIS model, each edge of an infected vertex transmits the epidemics with
rate $\lambda$, while infected nodes recover spontaneously with constant rate
$\mu$. The latter is fixed to $\mu=1$ without loss of generality. The model can
be simulated with the optimized Gillespie scheme proposed in
Ref.~\cite{Ferreira2012}. See also Ref.~\cite{Cota2017} for more details.

We consider quasi-stationary simulations~\cite{DeOliveira2005} in which the
dynamics returns to a previously  visited active configuration whenever the
absorbing state, consisting of all vertices susceptible, is visited. This
strategy permits to circumvent the difficulties of dealing with the
absorbing-state, which is the only true stationary state for any
finite-size networks. More details can be found
in Refs.~\cite{Cota2017,Sander2016}.

The effective transition point $\lambda_{c}(N)$, above which the epidemic
remains in an active phase for very long periods can be estimated using the
position of the maximum of the dynamical susceptibility~\cite{Ferreira2012}
\begin{equation}
\psi=N\frac{\langle \rho^{2}\rangle-\langle \rho \rangle}{\langle \rho \rangle}.
\label{eq:sus}
\end{equation}

The choice of structural (for $\gamma<3$) and rigid (for $\gamma>3$) upper
cutoffs allows the determination of the epidemic threshold unambiguously,
avoiding multiple peaks and the smearing of the transition that can appear for
SIS on power-law degree distribution networks, specially with large values of
$\gamma$~\cite{Ferreira2012,Mata2015}.

\subsection{Mean-field theories for correlated networks}
\label{sec:Mean_field}

In this subsection we summarize the predictions of the theoretical
approaches that will be compared to numerical simulations in Sec.~\ref{sec:results}.
For QMF and PQMF approaches, the equations for uncorrelated
and correlated networks are formally the same: correlations have only the
effect of modifying the entries of the adjacency matrix $A_{ij}$.

\subsubsection{Correlated Heterogeneous Mean-Field theory}
\label{sec:hmf}
HMF theory takes into account nearest-neighbors correlations
by the explicit consideration of the conditional probability $P(k'|k)$.
The HMF equation for the density of infected vertices with degree $k$, $\rho_{k}$, is
given by~\cite{Boguna2002}
\begin{equation}
\frac{d\rho_{k}}{dt}=-\rho_{k}+ (1-\rho_{k})\lambda \sum_{l}kP(l|k)\rho_{l},
\label{eq:hmf1}
\end{equation}
which yields an epidemic threshold given by
\begin{equation}
\lambda_{c}\Upsilon^{(1)}=1,
\label{eq:HMF_limiar}
\end{equation}
where $\Upsilon^{(1)}$ is the largest eigenvalue of the connectivity matrix $C_{kl}=kP(l|k)$.
For WPCM networks we have, $P(l|k)=P_\text{e}(l)f(l,k)$, therefore
\begin{equation}
C_{kl}=\frac{klP(l)}{\av{k}}f(l,k).
\label{eq:connectivity_matrix}
\end{equation}
In the absence of correlations, $C_{kl}=\frac{klP(l)}{\langle k\rangle}$,
implying that
$\lambda_{c}=\frac{\av{k}}{\av{k^2}}$~\cite{Boguna2002,Boguna2003}. It has been
shown~\cite{Boguna2003} that the HMF threshold vanishes for scale-free networks
with $2<\gamma<3$ in the thermodynamic limit, irrespective of degree
correlations.

\subsubsection{Quenched Mean-Field theory}
\label{subsec:qmf}

According to the QMF theory, which  neglects pairwise dynamical correlations, the
evolution of the probability $\rho_i$ that a vertex $i$ is infected is given
by~\cite{Chakrabarti2008}
\begin{equation}
\frac{d\rho_i}{d t} = -\rho_i+\lambda(1-\rho_i)\sum_{j=1}^{N}A_{ij}\rho_j,
\label{eq:qmf}
\end{equation}
where $N$ is the network size.
The epidemic threshold is given by
\begin{equation}
\lambda_\text{c}^\text{QMF} \Lambda^{(1)}=1
\end{equation}
where $\Lambda^{(1)}$ is the largest eigenvalue (LEV) of the adjacency matrix $A_{ij}$.
In the steady state we have
\begin{equation}
\rho_i = \frac{\lambda \sum_{j}A_{ij}\rho_j}{1+\lambda \sum_{j}A_{ij}\rho_j}.
\label{eq:rhoi_st1}
\end{equation}
Using Eq.~\eqref{eq:rhoi_st1}, Goltsev~\textit{et al}.~\cite{Goltsev2012}
have shown that $\rho_i\sim v^{(1)}_i$for  $\lambda\gtrsim \lambda_c^\text{QMF}$,
where $\lbrace v^{(1)}_i\rbrace$ is
the principal eigenvector (PEV)  corresponding to the LEV of $A_{ij}$, $\sum_{i}A_{ij}
v^{(1)}_j = \Lambda^{(1)} v^{(1)}_i$.
So, the order parameter
$\rho=\sum_i\rho_i/N$ of the QMF theory vanishes at $\lambda_c^\text{QMF}$ as
\begin{equation}
\rho \simeq a_{1} (\lambda\Lambda^{(1)}-1)
\label{eq:rho1}
\end{equation}
where
\begin{equation}
a_{1}(N)= \frac{\sum_{i=1}^{N} v^{(1)}_i}{N \sum_{i=1}^{N} \left[v^{(1)}_i\right]^3}.
\label{eq:alpha1}
\end{equation}
This same result was obtained independently in Ref.~\cite{VanMieghem2012}.

Within the QMF framework, Equation~(\ref{eq:rho1}) works well, close to
the threshold $\lambda_c^\text{QMF}$, under the hypothesis that the
network presents a spectral gap, i.e., the second largest eigenvalue
of $A_{ij}$ is much smaller than the first,
$\Lambda^{(1)} \gg \Lambda^{(2)}$ .
According to
Eqs.~\eqref{eq:rho1} and \eqref{eq:alpha1}, the QMF theory predicts the
existence of an endemic state, with a finite fraction of infected vertices above
the threshold $\lambda_c^\text{QMF}=1/\Lambda^{(1)}$, only if
$a_1\sim\mathcal{O}(1)$, which occurs when the PEV is delocalized. Localization
can be quantified by the inverse participation ratio (IPR) for the normalized
PEV~\cite{Goltsev2012}, defined as
\begin{equation}
Y_4=\sum_{i=1}^{N} \left[v^{(1)}_i\right]^4.
\label{eq:IPR}
\end{equation}
If the PEV is delocalized then $Y_4 \sim N^{-1}$, while $Y_4\sim \mathcal{O}(1)$
if the PEV is localized on a finite number of vertices, but weaker forms of
localization can be observed~\cite{Pastor-Satorras2015a}.

For random uncorrelated power-law networks the PEV is always
localized~\cite{Pastor-Satorras2015a}.  For $\gamma<5/2$ it is (weakly)
localized on a subextensive set of nodes coinciding with the maximum $K$-core, a
subgraph of strongly mutually interconnected nodes with degree larger than or
equal to $K$~\cite{Seidman1983269}. In such a case $Y_4 \sim N^{(\gamma-3)/2}$.
For $\gamma>5/2$ it is instead strongly localized on the largest hub plus its
nearest neighbors and $Y_4\sim \mathcal{O}(1)$~\cite{Pastor-Satorras2018}.
Hence within QMF theory the threshold separates the absorbing-phase from an
active but strictly nonendemic state.  However this does not imply that QMF
predictions are necessarily flawed.  Equation~(\ref{eq:qmf}) factorizes the
state of nearest neighbors and thus neglects dynamical correlations among them.
These dynamical correlations actually transmit the infection from the localized
PEV to the rest of the network, and thus may in principle transform the active
but localized state just above $\lambda_\text{c}^\text{QMF}$ into a full-fledged
endemic state~\cite{Boguna2013,Castellano2019}.

\subsubsection{Pair Quenched Mean-Field theory}
\label{subsec:pqmf}

An improvement with respect to QMF theory is obtained by taking into
account some dynamical correlations using the pairwise approximation
developed in Ref.~\cite{Mata2013}, where all derivation details can be
found.
Consider the probability $\phi_{ij}$ that a vertex $i$
  is susceptible and a neighbor $j$ is infected. The dynamical system
  to be solved is
\begin{equation}
\frac{d\rho_i}{dt} = -\rho_i+\lambda\sum_j\phi_{ij}A_{ij}
\label{eq:rhoi_pair}
\end{equation}
and
\begin{eqnarray}
\frac{d\phi_{ij}}{dt}   & = &  -(2+\lambda)\phi_{ij}+\rho_j
+\lambda\sum_{l} \frac{\omega_{ij}\phi_{jl}}{1-\rho_j}(A_{jl}-\delta_{il}) \nonumber \\
& - & \lambda\sum_{l} \frac{\phi_{ij}{\phi}_{il}}{1-\rho_i}(A_{il}-\delta_{lj}) ,
\label{eq:phi2}
\end{eqnarray}
where $\omega_{ij}=1-\phi_{ij}-\rho_i$.

Here, we develop a bit further the theory to analyze the steady state near the
critical point. Keeping only leading terms up to second order in $\rho_i$ in
Eq.~\eqref{eq:phi2} we obtain
\begin{equation}
\phi_{ij}\approx \frac{(2+\lambda)\rho_j-\lambda\rho_i}{2+2\lambda}-\rho_i\rho_j
+\mathcal{O}(\rho^3,\lambda\rho^2),
\label{eq:phi_qst}
\end{equation}
where we kept only leading order in $\lambda\approx \lambda_c\ll
1$~\cite{Mata2013} for quadratic terms in $\rho_i$.  Plugging
Eq.~\eqref{eq:phi_qst} in Eq.~\eqref{eq:rhoi_pair} with $d\rho_i/dt=0$,
we obtain
\begin{equation}
\rho_i = \frac{\lambda \sum_{i}B_{ij}(\lambda)\rho_j}{1+\lambda \sum_{i}B_{ij}(\lambda)\rho_j},
\label{eq:rhoi_st}
\end{equation}
where
\begin{equation}
B_{ij}=\frac{2+\lambda}{2\lambda+2}\frac{A_{ij}}{1+\frac{\lambda^{2}k_{i}}{2\lambda+2}}
\simeq \frac{A_{ij}}{1+\frac{\lambda^2 k_i}{2}},
\label{eq:Bij}
\end{equation}
is an effective, weighted adjacency matrix. The last passage in
Eq.~(\ref{eq:Bij}) assumes $\lambda \ll 1$.

Equation~\eqref{eq:rhoi_st} has exactly the same form of the stationary $\rho_i$
in Eq.~\eqref{eq:rhoi_st1}, obtained for QMF theory, replacing $A_{ij}$ by
$B_{ij}$. Therefore, all the spectral analysis described in
subsection~\ref{subsec:qmf} found for QMF theory can be extended to the PQMF
case with the replacement of spectral properties of $A_{ij}$ by those of
$B_{ij}$. For example, the epidemic threshold is given by
\begin{equation}
\lambda_\text{c}^\text{PQMF} \Omega^{(1)}(\lambda_\text{c}^\text{PQMF})=1
\end{equation}
where $\Omega^{(1)}$ is the largest eigenvalue of $B_{ij}$. One can
check that this result is exactly the same presented in
Ref.~\cite{Mata2013} expressed in a different way.  For
$\lambda\gtrsim \lambda_\text{c}^\text{PQMF}$ we have that $\rho_i
\sim w_i^{(1)}$, where $\lbrace w_i^{(1)}\rbrace$ is the PEV of
$B_{ij}(\lambda_\text{c}^\text{PQMF})$ and $\rho \simeq
b_1(\lambda\Omega^{(1)}(\lambda_\text{c}^\text{PQMF})-1)$ where
$b_1(N)$ has the same form of Eq.~\eqref{eq:alpha1} replacing $v_i$ by
$w_i$. So, the IPR of $\lbrace w_i^{(1)}\rbrace$, denoted by
$Y_4[B_{ij}]$, allows to quantify the localization in the PQMF theory.

\section{Results}
\label{sec:results}

\subsection{Accuracy of the theoretical estimates for the epidemic threshold}
\label{sec:accur-theor-estim}

\begin{figure}[ht]
  \centering
  \includegraphics[width=0.98\linewidth]{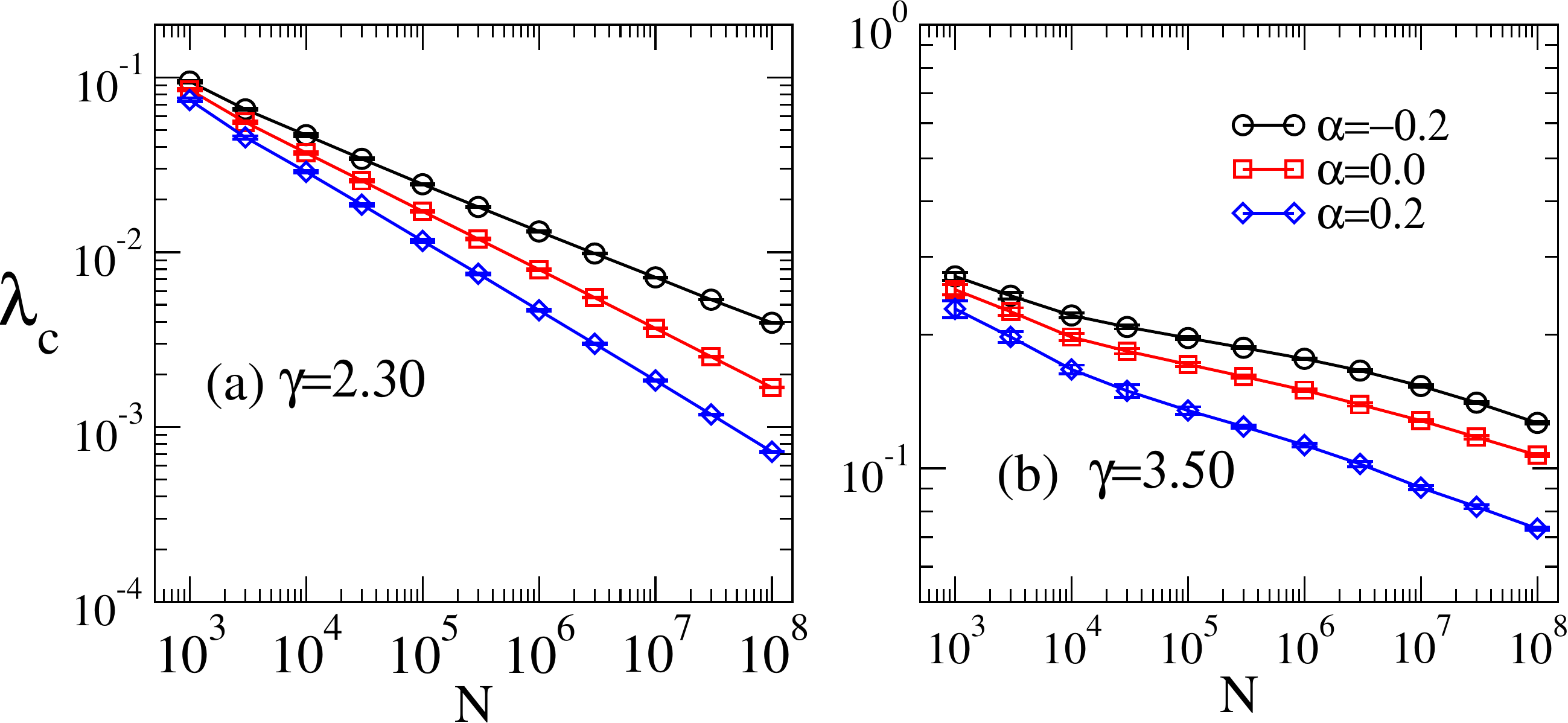}
  \caption{Threshold as a function of the network size for (a)
    $\gamma=2.3$, (b) $\gamma=3.5$ and different values of
    $\alpha$.  The lower cutoff is $\kmin=3$ for all curves
    while the upper cutoff is $\kmax=2\sqrt{N}$ for $\gamma<3$
    and $\kmax\sim N^{1/\gamma}$ for $\gamma>3$. Curves are
    averages over $10$ networks; error bars are smaller than
    symbols.}
  \label{fig:limiar1}
\end{figure}
Figure~\ref{fig:limiar1} shows the dependence of the epidemic threshold as a
function of the network size obtained in simulations with different values of
$\gamma$ and $\alpha$. We concentrate for the moment on two values of $\gamma$,
representative of the cases $\gamma<5/2$ and $\gamma>3$, for which the physical
mechanisms underlying the epidemic transition are clear~\cite{Castellano2012}.
Later we will discuss the case $5/2 < \gamma < 3$, whose interpretation is
hampered by extremely long crossover phenomena in the spectral properties.  As
we can see from this figure, all thresholds vanish as $N$ diverges, regardless
of the correlation level ($\alpha$) and heterogeneity ($\gamma$).  Compared to
the uncorrelated case, assortative networks ($\alpha>0$) have a smaller
threshold, while the threshold is larger for $\alpha<0$, i.e., disassortative
mixing, in agreement with the behavior of the LEV of the adjacency
matrix~\cite{PVM_assortativity_EJB2010,Goltsev2012}.  In the case $\gamma>3$,
this phenomenology can be qualitatively explained by considering the mechanism
of long-range mutual reinfection of
hubs~\cite{Boguna2013,Ferreira2016a,Castellano2019}, which triggers the epidemic
transition.  According to this mechanism, the subgraph consisting of the hub
plus its nearest-neighbors can sustain in isolation an active state for times
long enough to permit the activation of other hubs, even if they are not
directly connected.  This mechanism is at work independent of degree
correlations, as long as distances among hubs increase slowly enough with
network size.  In assortative networks, communication among hubs is enhanced
since they have larger probability to be closer; for disassortative topology the
converse is true and larger values of $\lambda$ are needed to trigger the
transition.

\begin{figure}[hbt]
  \centering
  \includegraphics[width=0.8\linewidth]{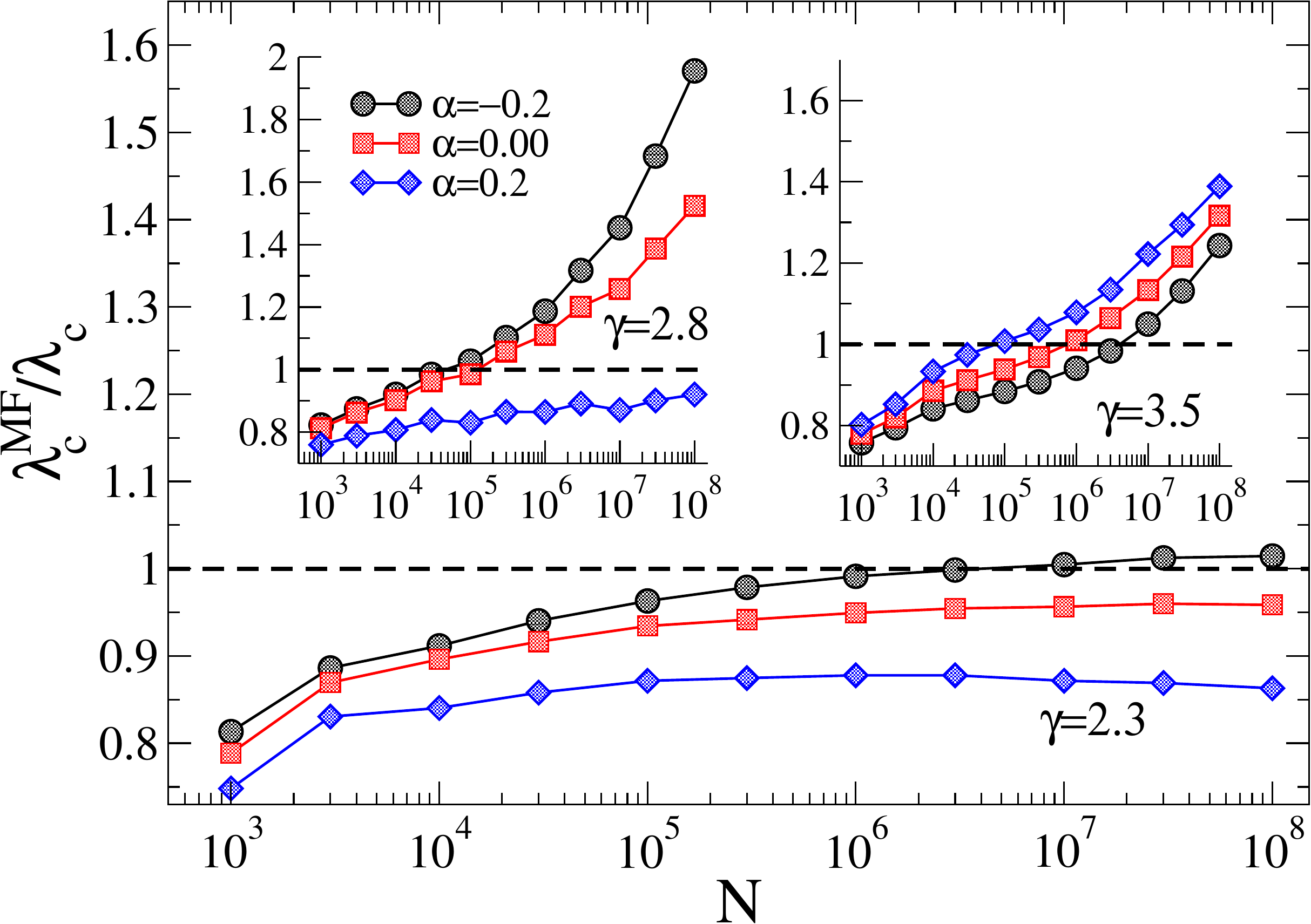}
  \caption{Ratio between thresholds of HMF theories
    ($\lambda_{c}^\text{MF}$) and simulations ($\lambda_\text{c}$) as a function
    of the network size for different values of $\gamma$ and
    $\alpha$. Main panel, right and left insets correspond to
    $\gamma=2.3$, 2.8, and 3.5 respectively. An upper cutoff
    $\kmax=2\sqrt{N}$ is considered for $\gamma<3$, while for
    $\gamma=3.5$, $\kmax\sim N^{1/\gamma}$. Averages correspond to 10
    network realizations and error bars are smaller than symbols.}
  \label{fig:hmf}
\end{figure}
The accuracy of HMF theory is tested with respect to simulations in
Fig.~\ref{fig:hmf}. For $\gamma=2.3$, we see a non-negligible asymptotic
discrepancy between HMF and simulations in the case of correlated networks.
Interestingly HMF {appears to overestimate the threshold} for disassortative
networks, while it underestimates it for assortative ones.  For larger values of
$\gamma$ the discrepancy is conspicuous and the epidemic threshold is
significantly overestimated, as can be seen in the insets of Fig.~\ref{fig:hmf}.

Comparisons between QMF and PQMF theories and simulations are shown in
Figs.~\ref{fig:spectral_vs_theory_g230}(a)
and~\ref{fig:spectral_vs_theory_g350}(a) in the range of network size
$10^3\le N\le10^8$.  For $\gamma=2.3$, both QMF and PQMF theories appear to
converge asymptotically to the epidemic threshold observed in simulations.  PQMF
displays a faster convergence than QMF, this effect being enhanced for smaller
values of $\alpha$.  For $\gamma=3.5$, the predictions of PQMF and QMF theories
succeed, qualitatively, in predicting that the threshold approaches zero in the
thermodynamic limit even in the presence of correlations.  However, the
theoretical threshold estimates depart from simulation results leading to
decreasing ratios $\lambda_{c}^{MF}/\lambda_{c}$ in the large network limit.  We
expect this ratio to decrease asymptotically as $1/\ln(\kmax)$
~\cite{Castellano2019}, in agreement with recent rigorous
results~\cite{Huang2018}.  Again, PQMF theory performs better than QMF.  In this
case, the improvement of PQMF over QMF grows with $\alpha$.


\subsection{Relation with spectral properties}

What is the origin of the discrepancies between theoretical predictions
and numerical results observed in Section~\ref{sec:accur-theor-estim}?

In this subsection we investigate which spectral feature is correlated with the
performance of the theoretical approaches depends.  We consider both QMF and
PQMF theories, testing their accuracy against the spectral properties of
adjacency matrices $A_{ij}$ and $B_{ij}$, respectively.

Let us consider first the case $\gamma=3.5$.  The real threshold is not the QMF
one because the PEV is localized.  As pointed out in
		Ref.~\cite{Goltsev2012} this in principle implies that the actual threshold
		coincides with the inverse of the largest eigenvalue corresponding to a
		delocalized PEV, coinciding with the HMF threshold
		$\lambda_c^\text{HMF}=\av{k}/\av{k^2}$.  But
actually the QMF approach neglects dynamical correlations, which have the effect
of allowing mutual reinfection events among different hubs in the network. In
this way an endemic global state can be established thanks to the long-range
interactions among localized states~\cite{Castellano2019} setting the actual
threshold to an intermediate value: $\lambda_c^{QMF} < \lambda_c <
\lambda_c^{HMF}$.  With this picture in mind we can predict that, if the
localization is stronger (higher values of the IPR $Y_4$), the actual threshold
will be farther from $\lambda_c^{QMF}$ and thus the performance (accuracy) of
the QMF approach will be reduced.

\begin{figure}[t]
  \includegraphics[width=0.95\linewidth]{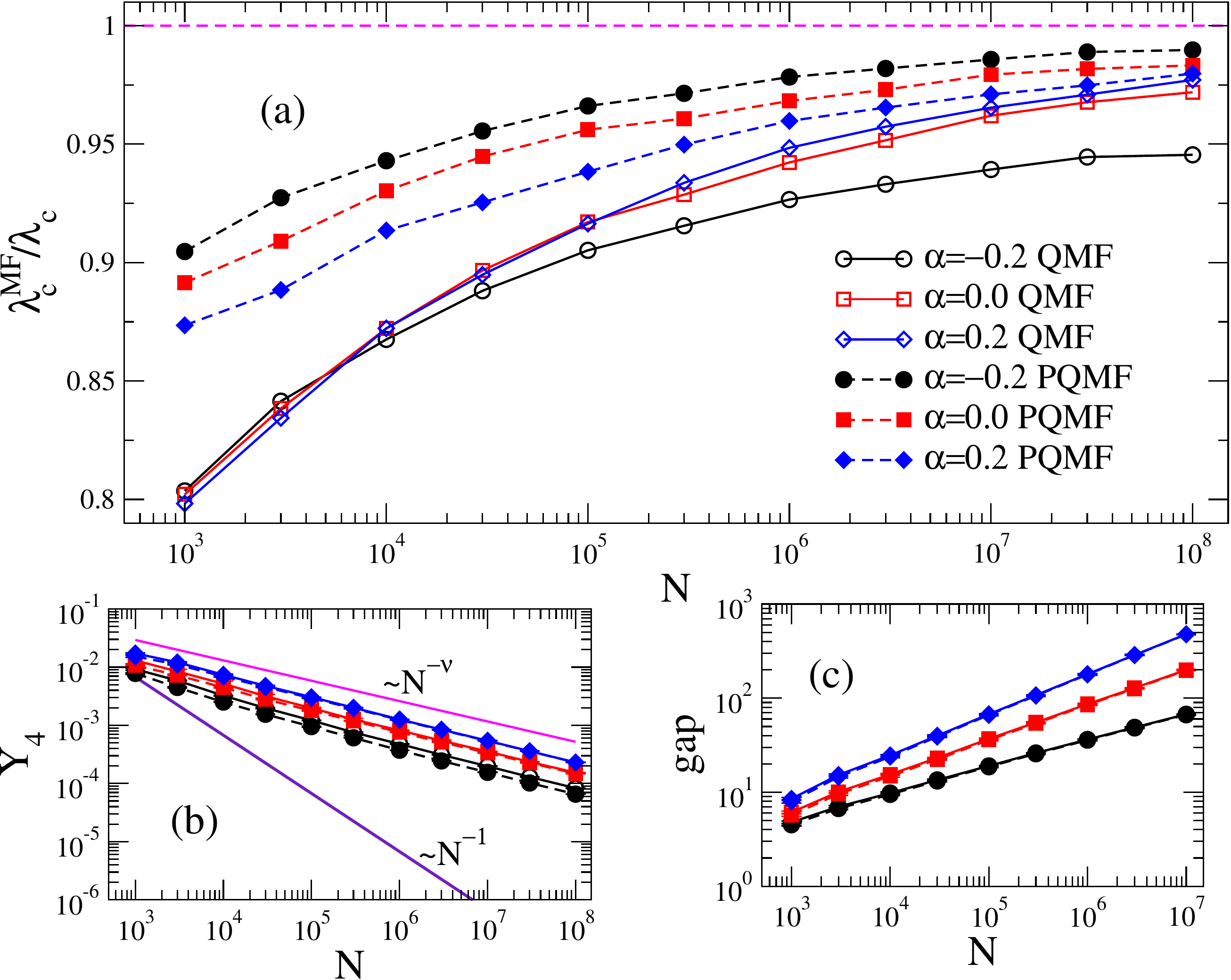}
  \caption{(a) Comparison of the QMF and PQMF mean-field theories, (b)
    IPR, and (c) spectral gap of $A_{ij}$ and $B_{ij}$ against size
    for $\gamma=2.3$ and different values of $\alpha$. Averages
    correspond to 10 network realizations. In (b), solid lines are
    power-law decays $Y_4\sim N^{-\nu}$ with $\nu=(3-\gamma)/2$ and
    $Y_4\sim N^{-1}$ corresponding to localization in the maximum
    $K$-core and finite set of vertices, respectively. Solid lines and
    empty symbols correspond to the QMF theory and $A_{ij}$ analysis
    while dashed lines and full symbols correspond to PQMF and
    critical $B_{ij}$.}
  \label{fig:spectral_vs_theory_g230}
\end{figure}

\begin{figure}[ht]
  \includegraphics[width=0.95\linewidth]{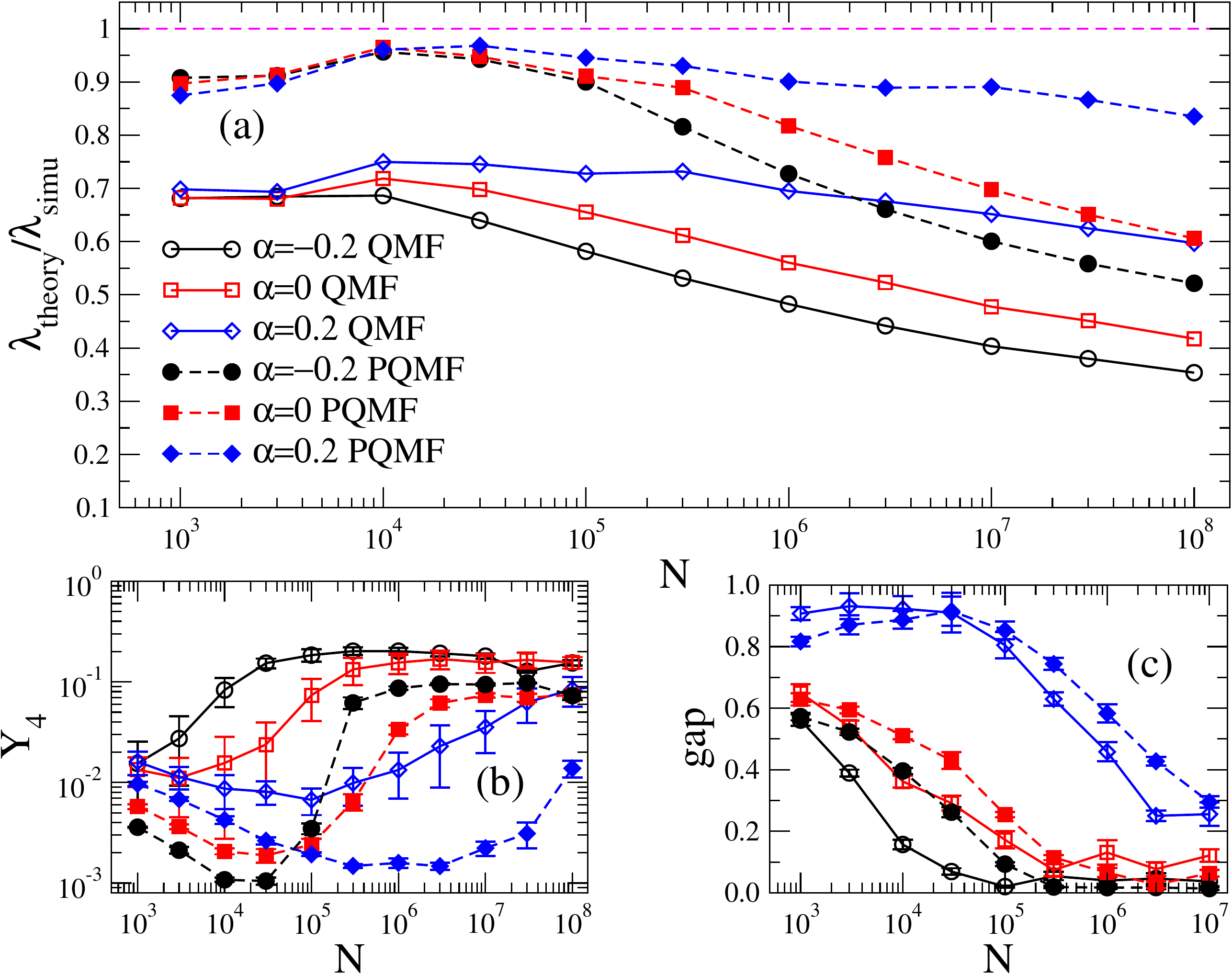}
  \caption{(a) Comparison of the QMF and PQMF mean-field theories, (b)
    IPR, and (c) spectral gap of $A_{ij}$ and $B_{ij}$ against size
    for $\gamma=3.5$ and different values of $\alpha$, using an upper
    cutoff $\kmax\sim N^{1/\gamma}$.  Averages correspond to 10
    network realizations. Solid lines and empty symbols correspond to
    the QMF theory and $A_{ij}$ analysis while dashed lines and full
    symbols correspond to PQMF and critical $B_{ij}$.}
  \label{fig:spectral_vs_theory_g350}
\end{figure}

We plot the dependence of $Y_4$ on the system size $N$ for $\gamma=3.5$ in
Fig.~\ref{fig:spectral_vs_theory_g350}(b):
The IPR of $A_{ij}$  converges to a finite value in the thermodynamic limit,
irrespective of the correlation degree, representing a PEV localized on a
finite set of vertices~\cite{Pastor-Satorras2015a,Goltsev2012}.
The saturation with size occurs earlier for disassortative and later for
assortative correlations, compared to the uncorrelated case.
In general, for a given size $N$, $Y_4$ is larger for smaller $\alpha$.
As expected, a better QMF performance occurs for smaller $Y_4$.

The IPR analysis for the PQMF theory, involving $B_{ij}$, has a qualitatively
similar behavior of QMF, but presenting lower values for the IPR. Hence, the
PQMF steady state solution is less localized than that of the QMF
theory. Correspondingly, the PQMF performance is better than the QMF
performance.  We also calculate, in Figure~\ref{fig:spectral_vs_theory_g350}(c),
the dependence of the spectral gap on the system size, both for the adjacency
matrix $A_{ij}$ (involved in QMF) and $B_{ij}$ (entering in PQMF).  The spectral
gap is defined as the difference $\Lambda^{(1)} - \Lambda^{(2)}$ between the
largest and second largest positive eigenvalues of the adjacency matrices.  The
gap of the adjacency matrix $A_{ij}$ is small and it decreases as $N$ grows, as
predicted by Ref.~\cite{Chung03}.  The gap is smaller for smaller $\alpha$.  The
dependence of spectral gap of $B_{ij}$ on size is qualitatively similar to the
gap of $A_{ij}$.

Notice that, while the amplitude of the spectral gap matters for the validity of
the QMF prediction for the prevalence above the critical point
[Eq.~(\ref{eq:rho1})], it does not play any role in the determination of
$\lambda_\text{c}^\text{QMF}$. Therefore there is no conceptual reason for expecting a
correlation between QMF performance and spectral gap size.  We find numerically
such a correlation in Fig.~\ref{fig:spectral_vs_theory_g350}, but we cannot
attribute a causal meaning to it.

Let us consider now $\gamma=2.3$.
In this case the physical mechanism underlying the epidemic transition is
different, as it does not involve the interaction between distant hubs,
rather the extension of activity from the max K-core to the rest of the
network. The connection between QMF performance and localization is
not easily predictable.

As shown in Fig.~\ref{fig:spectral_vs_theory_g230}(b), the IPR for
$\gamma=2.3$ follows a power-law $Y_{4}\sim N^{-\nu}$ with
$\nu\approx(3-\gamma)/2$, which corresponds to the IPR localized in
the maximum $K$-core of the network~\cite{Pastor-Satorras2015a}.
Correlations leave the scaling exponent unchanged, altering only the prefactor,
the smaller $\alpha$ the smaller the IPR.  This means that the PEV is still
localized on a sub-extensive fraction of nodes.
However, since $Y_4$ increases with $\alpha$, the PEV is more localized for
positive than for negative $\alpha$.
The same is true for the matrix $B_{ij}$ of the PQMF theory.
Interestingly, the effect on the performance of the theoretical approaches
is opposite.
QMF works better for larger $Y_4$, PQMF works better for smaller
$Y_4$. We have no simple interpretation for this result.

Figure~\ref{fig:spectral_vs_theory_g230}(c) shows the spectral gap for the WPCM
networks with $\gamma=2.3$. In this case the gap increases with network size
and it is smaller for smaller $\alpha$.  This is true also for the spectral gap
of PQMF.
Finally, let us observe that there is almost no difference between the spectral
properties of $A_{ij}$ and $B_{ij}$ for $\gamma=2.3$. This is indeed not
surprising for $\alpha=0$ since the term $\lambda^2 k_i$ in the denominator of
Eq.~(\ref{eq:Bij}) is asymptotically negligible, because
$\lambda_c^2 \kmax \sim \kmax^{2\gamma-5} \rightarrow 0$ as $N\rightarrow\infty$
for $\gamma<5/2$.

\subsection{The intermediate case $2.5<\gamma<3$}

\begin{figure}[thb]
  \centering
  \includegraphics[width=0.95\linewidth]{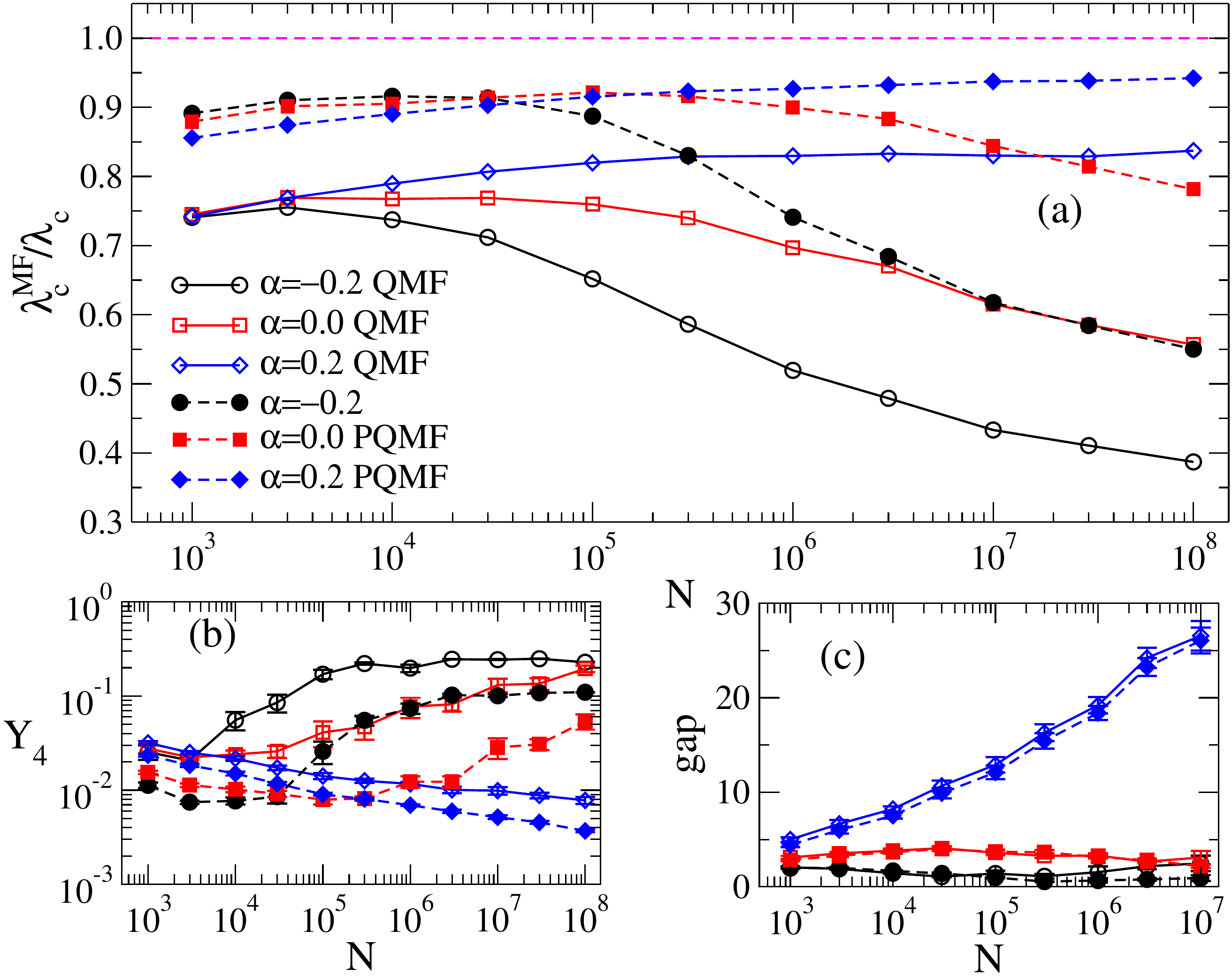}
  \caption{(a) Comparison of the QMF and PQMF mean-field theories, (b)
    IPR, and (c) spectral gap of $A_{ij}$ and $B_{ij}$ against size
    for $\gamma=2.8$ and different values of $\alpha$, using an upper
    cutoff $\kmax = 2\sqrt{N}$.  Averages correspond to 10 network
    realizations.  Solid lines and empty symbols correspond to the QMF
    theory and $A_{ij}$ analysis while dashed lines and full symbols
    correspond to PQMF and critical $B_{ij}$.}
	\label{fig:ucm_results_g280}
\end{figure}
As for the other values of $\gamma$, in this range the vanishing of the
threshold with $N$ is observed regardless of the correlation pattern.  The
localization phenomenon of the PEV in the case $5/2<\gamma<3$ is asymptotically
analogous to the case $\gamma>3$.  However, very strong crossover effects are
observed in this case, because of the presence of a localization process on the
max K-core (as for $\gamma<5/2$) competing with the localization around the
hub~\cite{Pastor-Satorras2015a}.  As a consequence, already in the uncorrelated
case, the PEV gets strongly localized around the largest hub only for very large
values of $N$.  Correlations further complicate the picture: Panel (b) of
Fig.~\ref{fig:ucm_results_g280} shows that disassortative correlations
accelerate the convergence to the final localized state.  For $\alpha>0$
instead, $Y_4$ is a decreasing function of $N$.  The upward bend of the curve
hints at an incipient crossover, but one cannot exclude that the asymptotic
behavior is different for $\alpha > 0$.  A similar pattern is observed for what
concerns the spectral gap (Fig.~\ref{fig:ucm_results_g280}(c)).

With regard to the performance of the theoretical approaches, for negative or
zero correlations the scenario perfectly matches what happens for $\gamma>3$:
all theories somehow fail in capturing the way the threshold vanishes, with PQMF
being less inaccurate than the others. In the case $\alpha=0.2$ numerical
results seem to suggest that both theories describe quite well how the
threshold changes with the system size. However, the large crossover effects
mentioned above do not allow to draw any firm conclusion.

We can summarize our findings by stating that the performance in predicting the
behavior of epidemic threshold of the QMF and PQMF theories on WPCM networks is
correlated with the size of the spectral gap and the IPR of PEV of the
respective $A_{ij}$ and $B_{ij}$ matrices that rule the prevalence near to the
transition point. A large spectral gap or a low IPR lead to a good performance
of the mean-field theories while the converse, small gap or large IPR, lead to
deviations from the theoretical predictions. QMF seems to be more correlated
with the spectral gap while PQMF with the IPR, at least in the regime where the
gap is significant and the theories are accurate.

\begin{figure*}[t]
  \centering
  \includegraphics[width=0.7\linewidth]{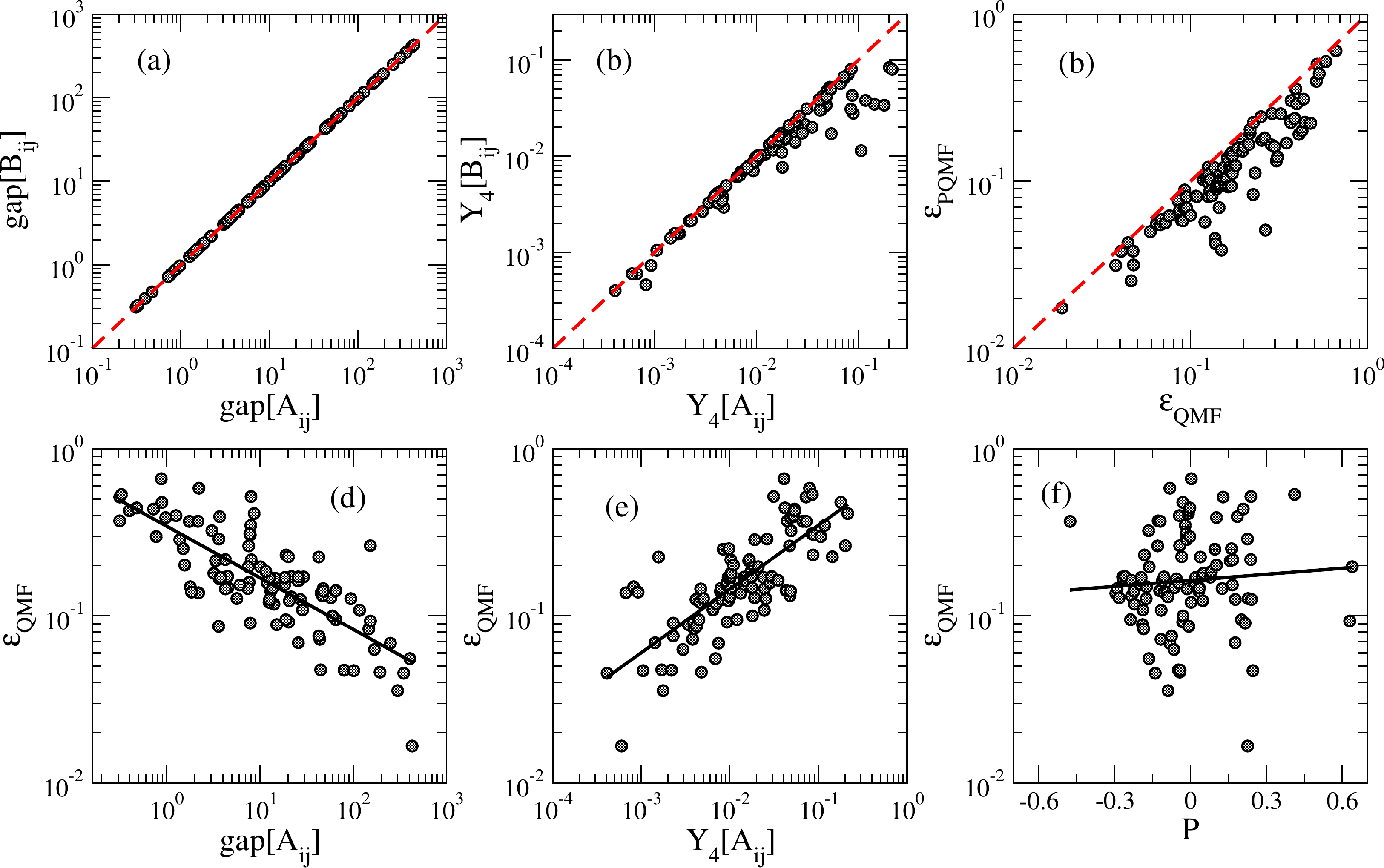}
  \caption{Scatter plots for a set of 99
    real networks (see Appendix~\ref{app:real}): each point corresponds
    to a single network. The
    spectral gap and the IPR of the matrix $B_{ij}$ is plotted
    versus the corresponding values for the matrix $A_{ij}$ in
    panels (a) and (b), respectively.
    In panel (c) we plot the relative errors
    of QMF and PQMF theoretical predictions with respect to the
    simulation, defined by Eq.~\eqref{eq:err}.
    Dashed red lines denote the diagonal. The relative errors of the
    QMF theory are plotted vs the spectral gap in panel (d), vs the
    IPR in panel (e), and vs the Pearson coefficient in panel (f).}
  \label{fig:real_net}
\end{figure*}

\subsection{Real networks}
\label{Real}
We extend our analysis to a set of 99 real-world networks encompassing a broad
range of origins, sizes and topological features, see Appendix~\ref{app:real}.
The spectral gap and IPR of matrices $A_{ij}$ and $B_{ij}$ are compared in the
scatter plots shown in Figs.~\ref{fig:real_net}(a) and (b). We see that the
spectral gap is almost the same for both adjacency matrices while the IPR
extracted from $B_{ij}$ is smaller than the one extracted from $A_{ij}$, in
particular in the range of large IPR values. This shows that the PQMF matrices
$B_{ij}$ are less localized than the matrix $A_{ij}$, relevant for QMF
theory. The relative errors between QMF or PQMF mean-field theories and
simulations, defined as
\begin{equation}
\varepsilon = \frac{\lambda_c-\lambda_c^\text{MF}}{\lambda_c},
\label{eq:err}
\end{equation}
are compared in the scatter plot shown in Fig.~\ref{fig:real_net}(c).
As in the case
random networks, PQMF outperforms QMF theory for all investigated networks.

On this set of networks, we test the relation observed for synthetic
correlated networks, connecting qualitatively the accuracy of QMF
and PQMF threshold predictions with the properties of the adjacency
matrices (spectral gap and IPR), respectively, and with the
Pearson coefficient $P$, measuring network topological correlations.
$P$ is defined as~\cite{Newman2014}
\begin{equation}
P = \frac{\sum_{ij}\left(A_{ij}-\frac{k_ik_j}{N\av{k}}\right)k_ik_j}
{\sum_{ij}\left(k_i\delta_{ij}-\frac{k_ik_j}{N\av{k}}\right)k_ik_j}.
\end{equation}
The Pearson coefficient lays in the interval $-1<P<1$, being negative for
disassortative, null for uncorrelated, and positive for assortative
networks. The analyses for QMF are shown in the scatter plots of the relative
error $\varepsilon$ against the corresponding topological properties in
Figs.~\ref{fig:real_net}(d)-(e).  Qualitatively similar patterns obtained for
PQMF are not shown.  We can see that in real networks, the correlation between
the performance of the theoretical prediction and the spectral gap is on average
the same as the one observed for the WPCM: A larger spectral gap is associated
to a larger accuracy.  The inverse correlation with the IPR is again preserved:
A smaller $Y_4$ corresponds to a more accurate prediction. We do not find
instead a significant correlation with the Pearson coefficient.  Statistical
analyses were performed using the correlation coefficients obtained from either
power-law, in the case of spectral gap and IPR, or from exponential, in the case
of Pearson coefficient, regressions of the scatter plots.
We obtain strong statistical correlations with $|r|\gtrsim 0.70$ (p-value
$<10^{-5}$) for both QMF and PQMF using either IPR or spectral gap of the
corresponding matrices. Values $r\lesssim 0.2$ (p-value $>0.05$) for correlation
with the Pearson coefficient of the network confirms no significant statistical
correlations.

\subsection{Epidemic prevalence near to the epidemic threshold}

\begin{figure}[hbt]
  \includegraphics[width=0.8\linewidth]{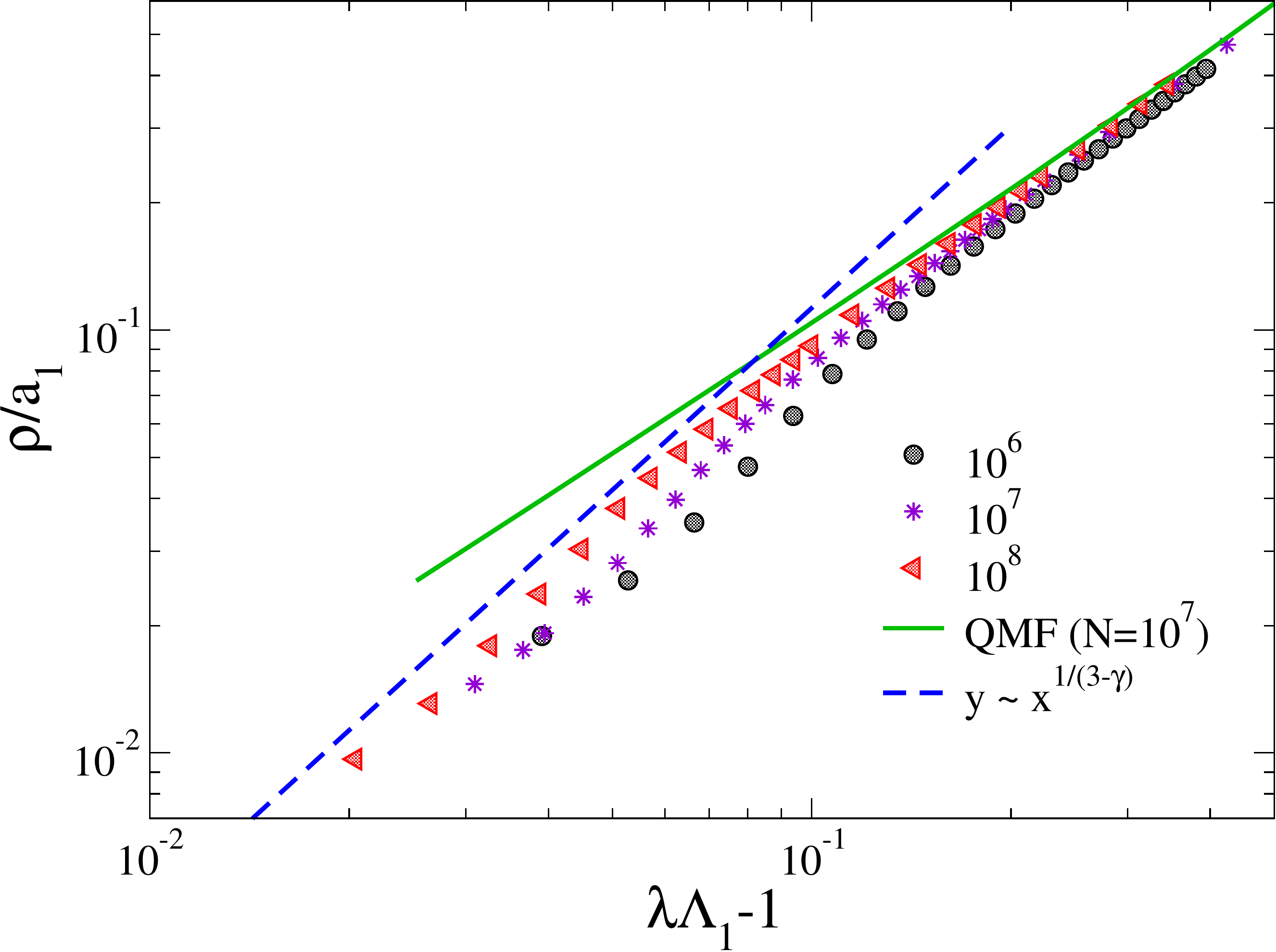}
  \caption{Rescaled average density as a function of the distance from the
  	epidemic threshold. QS simulations for different sizes indicated in the
  	legends. Solid line is a numerical integration of the QMF theory,
  	Eq.(\ref{eq:qmf}), for $N=10^7$ while the dashed one is a power law with
  	exponent predicted analytically in Ref.~\cite{mountford2013}. We used
  	uncorrelated networks ($\alpha=0$) with degree exponent $\gamma=2.3$ and
  	$\kmax=2\sqrt{N}$.}
  \label{fig:beta}
\end{figure}

Figure~\ref{fig:limiar1} shows that the QMF prediction for the epidemic
threshold tends to the same limit of numerical simulations both for uncorrelated
and correlated networks for $\gamma=2.3$.  This observation naturally leads to
wonder whether QMF is asymptotically an exact description for SIS dynamics on
random networks with $\gamma<5/2$.  In order to answer this question we test the
exactness of the other prediction of the QMF theory, Eq.~(\ref{eq:rho1}),
stating that the fraction of infected individuals decays to zero linearly as the
threshold is approached from above.  Numerical results, for the case of
uncorrelated networks $\alpha=0$, are shown in Fig.~\ref{fig:beta}, where the
density and the infection rates are rescaled to conform to Eq.~(\ref{eq:rho1}).
We can clearly see the existence of two scaling regimes.  For
$\lambda\Lambda_1-1\ll 1$ the density scales with an exponent larger than the
prediction $\beta^\text{(QMF)}=1$. The observed exponent is consistent with the
exact result of Ref.~\cite{mountford2013} $\beta=1/(3-\gamma)$ which is also
(probably accidentally) the value predicted by HMF
theory~\cite{Pastor-Satorras2002}.  This exponent is observed in a regime very
close to the transition, where the system is kept asymptotically active only by
virtue of the QS method. We performed a non-perturbative analysis by integrating
the QMF equations using a fourth order Runge-Kutta method for
$\lambda>\frac{1}{\Lambda^{(1)}}$ for $N=10^7$.
A comparison with simulation results confirms that the QMF theory correctly
predicts the linear behavior of the prevalence $\rho$ around the epidemic
transition, but only sufficiently far from it.  In the immediate neighborhood of
the threshold the decay is more rapid.

\section{Conclusions}
\label{sec:conclusion}

The determination of the epidemic threshold in models of disease spreading in
complex topologies is a nontrivial problem in network science. Several
theoretical approaches have been proposed, applying approximations with
different levels of stringency, that provide contrasting predictions on the
epidemic threshold.  Among the main theoretical approaches at the mean-field
level we can consider, in decreasing order of approximation, the heterogeneous
mean-field theory (HMF), neglecting dynamical correlations and the actual
pattern of connections in the network (preserving only its statistical
properties); the quenched mean-field theory (QMF), also neglecting dynamical
correlations but keeping the network structure; and the pair quenched mean-field
theory (PQMF) that incorporates dynamical correlations between pairs of
connected nodes.  In this paper we have presented a comparison of the
predictions of these three approximate theories for the case of the
susceptible-infected-susceptible (SIS) epidemic model, focusing on the case of
networks with a power-law degree distribution and degree correlations,
representative of many real networked systems.

Comparing the predictions with actual stochastic simulations of the SIS process,
we observe that, independently of the degree of correlations, the predictions of
PQMF are more accurate than those of QMF, while both outperform HMF, which fails
to predict the vanishing threshold observed for a degree exponent $\gamma > 3$.
While overall PQMF is more accurate than QMF, the two approximations show
different levels of accuracy when compared in networks with different levels of
correlations. Thus, for the case of synthetic networks generated with the
Weber-Porto algorithm~\cite{Weber2007}, we observe that, for fixed network size
and degree heterogeneity, QMF predictions are more accurate in assortative
networks than in disassortative ones. On the other hand, PQMF is increasingly
accurate in the presence of disassortative correlations for small degree
exponent, while it is more accurate when correlations are assortative if the
degree exponent is large.

We propose a criterion for the accuracy of the QMF and PQMF approaches based on
the spectral properties of the networks. We observe that the accuracy is
positively correlated with the amplitude of the spectral gap of the adjacency
matrix and is inversely related with degree of localization of the principal
eigenvalue, as measured by the inverse participation ratio.  This general
observation is corroborated by the analysis of a large set of real correlated
networks, covering a wide range of sizes and topological features.

Additionally, we investigate the behavior of the order parameter of the
transition, measured in terms of the prevalence or density of infected nodes in
the steady state, for $\gamma<5/2$. We observe that, in uncorrelated synthetic networks, the
linear decay (critical exponent $\beta=1$) predicted by QMF theory is observed
in stochastic simulations not very close to the transition.  When fluctuation
effects become more important, i.e., very close to the transition, the observed
exponent $\beta$ crosses over to the value $\beta = 1/(3-\gamma)$, in agreement
with rigorous mathematical results~\cite{mountford2013}.

\begin{acknowledgments}
  This work was partially supported by the Brazilian agencies CAPES,
  CNPq and FAPEMIG. S.C.F. thanks the support from the program
  \textit{Ci\^encia sem Fronteiras} - CAPES under project
  No. 88881.030375/2013-01. R.P.-S. and C.C. acknowledge financial
  support from the Spanish MINECO, under project FIS2016-76830-C2-1-P.
  R. P.-S. acknowledges additional financial support from ICREA
  Academia, funded by the Generalitat de Catalunya.  This study was
  financed in part by the Coordenação de Aperfeiçoamento de Pessoal de
  Nível Superior - Brasil (CAPES) - Finance Code 001.
\end{acknowledgments}

\appendix

\section{Summary of real networks properties}
\label{app:real}

We consider 99 real networks with diverse structural properties, based on
the lists of Refs. ~\cite{radicchi2015predicting,radicchi2015breaking}.
Here we investigate their giant connected components, after symmetrizing
all edges (weighted and/or directed) and avoiding multiple and self
connections.
The list of networks with some topological properties and epidemic (SIS)
parameters is shown in Table~\ref{tab:realnetsprops}. For detailed
information about the original references for all the networks, please
check Refs.~\cite{radicchi2015predicting,radicchi2015breaking}.

\begin{widetext}
\begin{center}
  \renewcommand*{\arraystretch}{1.3}

  \setlength{\LTcapwidth}{0.8\linewidth}
  \begin{longtable}{@{\extracolsep{\fill}}lcccccccccc@{}}
\caption{Properties of the set with 99 networks of distinct types.
We show the network size $N$, the average degree $\av{k}$, the Pearson
coefficient $P$, the IPRs of both $A_{ij}$ and critical $B_{ij}$ matrices, the
spectral gap of $A_{ij}$, the thresholds of  simulations ($\lambda_c$), QMF
($\lambda_c^\text{QMF}$) and PQMF ($\lambda_c^\text{PQMF}$) theories.}
\label{tab:realnetsprops}
\\
\hline\hline
Network & $N$ & $\av{k}$ & $P$ & IPR$_\text{A}$ & IPR$_\text{B}$ & $\Delta \Lambda^{1,2}_\text{A}$ & $\lambda_c$ &  $\lambda_c^\text{QMF}$ &$\lambda_c^\text{PQMF}$ \\
\hline
Karate club                  &  $34$             &  $4.59$      &  $-0.476$        &  $0.0730$        &  $0.0649$        &  $1.75$        &  $0.235$       &  $0.149$         &  $0.181$      \\    
Radoslaw Email               &  $167$            &  $38.9$      &  $-0.295$        &  $0.0133$        &  $0.0132$        &  $45.2$        &  $0.0191$      &  $0.0165$        &  $0.0168$     \\    
Spanish B                    &  $12,643$         &  $8.70$      &  $-0.290$        &  $0.0246$        &  $0.0174$        &  $47.5$        &  $0.0105$      &  $0.00897$       &  $0.00951$    \\    
Spanish A                    &  $11,558$         &  $7.45$      &  $-0.282$        &  $0.0190$        &  $0.0150$        &  $57.5$        &  $0.0113$      &  $0.00985$       &  $0.0103$     \\    
US Air Transportation        &  $500$            &  $11.9$      &  $-0.268$        &  $0.0176$        &  $0.0173$        &  $29.3$        &  $0.0251$      &  $0.0208$        &  $0.0214$     \\    
Little Rock Lake             &  $183$            &  $26.6$      &  $-0.266$        &  $0.0148$        &  $0.0145$        &  $14.6$        &  $0.0291$      &  $0.0242$        &  $0.0249$     \\    
Japanese                     &  $2,698$          &  $5.93$      &  $-0.259$        &  $0.0296$        &  $0.0214$        &  $21.7$        &  $0.0281$      &  $0.0233$        &  $0.0250$     \\    
English                      &  $7,377$          &  $12.0$      &  $-0.237$        &  $0.0120$        &  $0.0103$        &  $65.3$        &  $0.0101$      &  $0.00914$       &  $0.0094$     \\    
French                       &  $8,308$          &  $5.74$      &  $-0.233$        &  $0.0351$        &  $0.0200$        &  $26.3$        &  $0.0197$      &  $0.0165$        &  $0.0179$     \\    
Jung                         &  $6,120$          &  $16.4$      &  $-0.233$        &  $0.0478$        &  $0.0335$        &  $46.9$        &  $0.00810$     &  $0.00703$       &  $0.00743$    \\    
JDK                          &  $6,434$          &  $16.7$      &  $-0.223$        &  $0.0484$        &  $0.0341$        &  $47.6$        &  $0.00810$     &  $0.00696$       &  $0.00737$    \\    
Political blogs              &  $1,222$          &  $27.4$      &  $-0.221$        &  $0.00701$       &  $0.00687$       &  $14.1$        &  $0.0153$      &  $0.0135$        &  $0.0137$     \\    
Internet                     &  $22,963$         &  $4.22$      &  $-0.198$        &  $0.0146$        &  $0.0116$        &  $18.4$        &  $0.0165$      &  $0.0140$        &  $0.0148$     \\    
AS Caida                     &  $26,475$         &  $4.03$      &  $-0.195$        &  $0.0240$        &  $0.0140$        &  $18.5$        &  $0.0173$      &  $0.0144$        &  $0.0157$     \\    
EU email                     &  $224,832$        &  $3.02$      &  $-0.189$        &  $0.00340$       &  $0.00328$       &  $15.2$        &  $0.0107$      &  $0.00975$       &  $0.0101$     \\    
UC Irvine                    &  $1,893$          &  $14.6$      &  $-0.188$        &  $0.00643$       &  $0.00608$       &  $28.6$        &  $0.0233$      &  $0.0208$        &  $0.0214$     \\    
Linux, mailing list          &  $24,567$         &  $12.9$      &  $-0.185$        &  $0.00395$       &  $0.00386$       &  $147$         &  $0.00490$     &  $0.00448$       &  $0.00452$    \\    
AS Oregon                    &  $6,474$          &  $3.88$      &  $-0.182$        &  $0.0868$        &  $0.0429$        &  $19.1$        &  $0.0281$      &  $0.0216$        &  $0.0249$     \\    
Linux, soft.                 &  $30,817$         &  $13.8$      &  $-0.175$        &  $0.0256$        &  $0.0197$        &  $94.1$        &  $0.00670$     &  $0.00585$       &  $0.00616$    \\    
Gnutella, Aug. 25, 2002      &  $22,663$         &  $4.83$      &  $-0.173$        &  $0.000815$      &  $0.000464$      &  $1.79$        &  $0.108$       &  $0.0916$        &  $0.104$      \\    
Les Miserables               &  $77$             &  $6.60$      &  $-0.165$        &  $0.0492$        &  $0.0482$        &  $3.05$        &  $0.123$       &  $0.0833$        &  $0.0919$     \\    
Petster-cats                 &  $148,826$        &  $73.2$      &  $-0.164$        &  $0.00687$       &  $0.00635$       &  $405$         &  $0.000900$    &  $0.000847$      &  $0.000855$   \\    
C. Elegans, neural           &  $297$            &  $14.5$      &  $-0.163$        &  $0.0189$        &  $0.0176$        &  $10.1$        &  $0.0511$      &  $0.0410$        &  $0.0434$     \\    
Libimseti                    &  $220,970$        &  $156$       &  $-0.139$        &  $0.000406$      &  $0.000398$      &  $348$         &  $0.00110$     &  $0.00106$       &  $0.00106$    \\    
David Copperfield            &  $112$            &  $7.59$      &  $-0.129$        &  $0.0473$        &  $0.0397$        &  $7.57$        &  $0.103$       &  $0.0760$        &  $0.0844$     \\    
Political books              &  $105$            &  $8.40$      &  $-0.128$        &  $0.0444$        &  $0.0419$        &  $0.313$       &  $0.133$       &  $0.0838$        &  $0.0927$     \\    
Google                       &  $15,763$         &  $18.9$      &  $-0.122$        &  $0.0430$        &  $0.0303$        &  $65.1$        &  $0.00670$     &  $0.00575$       &  $0.00608$    \\    
Social 3                     &  $32$             &  $5.00$      &  $-0.119$        &  $0.0665$        &  $0.0568$        &  $2.16$        &  $0.265$       &  $0.167$         &  $0.205$      \\    
Euron                        &  $33,696$         &  $10.7$      &  $-0.116$        &  $0.00379$       &  $0.00361$       &  $43.9$        &  $0.00910$     &  $0.00844$       &  $0.00859$    \\    
Web Stanford                 &  $255,265$        &  $15.2$      &  $-0.116$        &  $0.0245$        &  $0.0230$        &  $117$         &  $0.00250$     &  $0.00223$       &  $0.0023$     \\    
Bay Wet                      &  $128$            &  $32.4$      &  $-0.112$        &  $0.0151$        &  $0.0147$        &  $25.6$        &  $0.0301$      &  $0.0252$        &  $0.0259$     \\    
Bay Dry                      &  $128$            &  $32.9$      &  $-0.104$        &  $0.0148$        &  $0.0145$        &  $25.7$        &  $0.0301$      &  $0.0249$        &  $0.0256$     \\    
Gnutella, Aug. 30, 2002      &  $36,646$         &  $4.82$      &  $-0.104$        &  $0.000672$      &  $0.000604$      &  $2.19$        &  $0.0897$      &  $0.0773$        &  $0.0856$     \\    
Gnutella, Aug. 31, 2002      &  $62,561$         &  $4.73$      &  $-0.0927$       &  $0.000921$      &  $0.000731$      &  $1.85$        &  $0.0881$      &  $0.0759$        &  $0.0844$     \\    
Petster-hamster              &  $1,788$          &  $14.0$      &  $-0.0889$       &  $0.0100$        &  $0.00938$       &  $21.6$        &  $0.0249$      &  $0.0217$        &  $0.0223$     \\    
Petster-dogs                 &  $426,485$        &  $40.1$      &  $-0.0884$       &  $0.00176$       &  $0.00157$       &  $300$         &  $0.00140$     &  $0.00135$       &  $0.00136$    \\    
Network Science              &  $379$            &  $4.82$      &  $-0.0817$       &  $0.0794$        &  $0.0705$        &  $2.21$        &  $0.230$       &  $0.0964$        &  $0.110$      \\    
AS Skitter                   &  $1,694,616$      &  $13.1$      &  $-0.0814$       &  $0.00746$       &  $0.00709$       &  $251$         &  $0.00160$     &  $0.00149$       &  $0.00151$    \\    
Slashdot zoo                 &  $79,116$         &  $11.8$      &  $-0.0746$       &  $0.00229$       &  $0.00218$       &  $42.8$        &  $0.00830$     &  $0.00767$       &  $0.00778$    \\    
Wikipedia, edits             &  $113,123$        &  $35.8$      &  $-0.0651$       &  $0.00295$       &  $0.00266$       &  $169$         &  $0.0027$      &  $0.00253$       &  $0.00255$    \\    
CiteSeer                     &  $365,154$        &  $9.43$      &  $-0.0632$       &  $0.0177$        &  $0.0110$        &  $4.54$        &  $0.0202$      &  $0.0172$        &  $0.0183$     \\    
Cora                         &  $23,166$         &  $7.70$      &  $-0.0553$       &  $0.0100$        &  $0.00898$       &  $3.66$        &  $0.0381$      &  $0.0317$        &  $0.0334$     \\    
Thesaurus                    &  $23,132$         &  $25.7$      &  $-0.0477$       &  $0.0017$        &  $0.00156$       &  $44.7$        &  $0.0105$      &  $0.0100$        &  $0.0102$     \\    
DBLP, citations              &  $12,495$         &  $7.93$      &  $-0.0461$       &  $0.0282$        &  $0.0174$        &  $12.0$        &  $0.0277$      &  $0.0234$        &  $0.0251$     \\    
Dolphins                     &  $62$             &  $5.13$      &  $-0.0436$       &  $0.0526$        &  $0.0493$        &  $1.26$        &  $0.231$       &  $0.139$         &  $0.164$      \\    
DBpedia                      &  $3,915,921$      &  $6.42$      &  $-0.0427$       &  $0.201$         &  $0.0840$        &  $153$         &  $0.00190$     &  $0.00140$       &  $0.00180$    \\    
Wikipedia, pages             &  $2,070,367$      &  $40.9$      &  $-0.0418$       &  $0.00477$       &  $0.00294$       &  $194$         &  $0.00130$     &  $0.00124$       &  $0.00127$    \\    
Epinions                     &  $75,877$         &  $10.7$      &  $-0.0406$       &  $0.00219$       &  $0.00211$       &  $79.8$        &  $0.00570$     &  $0.00543$       &  $0.00548$    \\    
Slashdot                     &  $51,083$         &  $4.56$      &  $-0.0347$       &  $0.144$         &  $0.0347$        &  $20.2$        &  $0.0219$      &  $0.0170$        &  $0.0201$     \\   
Hep-Th, citations            &  $27,400$         &  $25.7$      &  $-0.0305$       &  $0.00931$       &  $0.00705$       &  $20.3$        &  $0.00990$     &  $0.00899$       &  $0.00922$    \\    
S 838                        &  $512$            &  $3.20$      &  $-0.0300$       &  $0.179$         &  $0.0340$        &  $0.889$       &  $0.382$       &  $0.200$         &  $0.297$      \\   
Gowalla                      &  $196,591$        &  $9.67$      &  $-0.0293$       &  $0.0180$        &  $0.00764$       &  $60.0$        &  $0.00650$     &  $0.00585$       &  $0.00609$    \\    
Amazon, Mar. 12, 2003        &  $400,727$        &  $11.7$      &  $-0.0203$       &  $0.118$         &  $0.0381$        &  $7.97$        &  $0.0273$      &  $0.0178$        &  $0.0227$     \\    
Amazon, Jun. 6, 2003         &  $403,364$        &  $12.1$      &  $-0.0176$       &  $0.0891$        &  $0.0279$        &  $7.87$        &  $0.0252$      &  $0.0175$        &  $0.0219$     \\    
Amazon, May. 5, 2003         &  $410,236$        &  $11.9$      &  $-0.0169$       &  $0.0843$        &  $0.0309$        &  $7.79$        &  $0.0249$      &  $0.0172$        &  $0.0214$     \\    
Air traffic                  &  $1,226$          &  $3.93$      &  $-0.0152$       &  $0.0191$        &  $0.0154$        &  $1.38$        &  $0.152$       &  $0.109$         &  $0.127$      \\    
Gnutella, Aug. 4, 2002       &  $10,876$         &  $7.35$      &  $-0.0132$       &  $0.00469$       &  $0.00377$       &  $4.26$        &  $0.0685$      &  $0.0586$        &  $0.0637$     \\    
Gnutella, Aug. 24, 2002      &  $26,498$         &  $4.93$      &  $-0.00778$      &  $0.214$         &  $0.0800$        &  $8.68$        &  $0.0865$      &  $0.0511$        &  $0.0700$     \\    
Hep-Ph, citations            &  $34,401$         &  $24.5$      &  $-0.00644$      &  $0.00421$       &  $0.00323$       &  $3.61$        &  $0.0143$      &  $0.0131$        &  $0.0133$     \\    
S 420                        &  $252$            &  $3.17$      &  $-0.00591$      &  $0.0542$        &  $0.0172$        &  $0.398$       &  $0.400$       &  $0.229$         &  $0.320$      \\    
Amazon, May. 2, 2003         &  $262,111$        &  $6.87$      &  $-0.00248$      &  $0.106$         &  $0.0114$        &  $0.771$       &  $0.0605$      &  $0.0425$        &  $0.0508$     \\    
S 208                        &  $122$            &  $3.10$      &  $-0.00201$      &  $0.0419$        &  $0.0301$        &  $0.475$       &  $0.437$       &  $0.244$         &  $0.338$      \\    
Digg                         &  $29,652$         &  $5.72$      &  $0.00265$       &  $0.00457$       &  $0.00346$       &  $12.5$        &  $0.0369$      &  $0.0324$        &  $0.0348$     \\    
US Power grid                &  $4,941$          &  $2.67$      &  $0.00346$       &  $0.0409$        &  $0.0386$        &  $0.874$       &  $0.396$       &  $0.134$         &  $0.157$      \\    
Gnutella, Aug. 5, 2002       &  $8,842$          &  $7.20$      &  $0.0146$        &  $0.00931$       &  $0.00855$       &  $7.68$        &  $0.0505$      &  $0.0425$        &  $0.0453$     \\    
Jazz                         &  $198$            &  $27.7$      &  $0.0202$        &  $0.0143$        &  $0.0141$        &  $12.6$        &  $0.0301$      &  $0.0250$        &  $0.0257$     \\    
Gnutella, Aug. 9, 2002       &  $8,104$          &  $6.42$      &  $0.0331$        &  $0.00782$       &  $0.00737$       &  $13.1$        &  $0.0409$      &  $0.0351$        &  $0.0370$     \\    
Gnutella, Aug. 8, 2002       &  $6,299$          &  $6.60$      &  $0.0355$        &  $0.00795$       &  $0.00752$       &  $13.7$        &  $0.0413$      &  $0.0352$        &  $0.0371$     \\    
LiveJournal                  &  $5,189,808$      &  $18.8$      &  $0.0394$        &  $0.00157$       &  $0.00157$       &  $42.7$        &  $0.00240$     &  $0.00186$       &  $0.00186$    \\    
High school, 2012            &  $180$            &  $24.7$      &  $0.0464$        &  $0.0102$        &  $0.0101$        &  $4.49$        &  $0.0401$      &  $0.0332$        &  $0.0344$     \\    
Open flights                 &  $2,905$          &  $10.8$      &  $0.0489$        &  $0.00963$       &  $0.00942$       &  $20.6$        &  $0.0181$      &  $0.0159$        &  $0.0162$     \\    
Gnutella, Aug. 6, 2002       &  $8,717$          &  $7.23$      &  $0.0516$        &  $0.0103$        &  $0.00957$       &  $3.20$        &  $0.0545$      &  $0.0447$        &  $0.0478$     \\    
URV email                    &  $1,133$          &  $9.62$      &  $0.0782$        &  $0.00956$       &  $0.00865$       &  $3.78$        &  $0.0581$      &  $0.0482$        &  $0.0512$     \\    
High school, 2011            &  $126$            &  $27.1$      &  $0.0829$        &  $0.0173$        &  $0.0171$        &  $11.6$        &  $0.0361$      &  $0.0294$        &  $0.0304$     \\    
DBLP, collaborations         &  $1,137,114$      &  $8.83$      &  $0.0964$        &  $0.00797$       &  $0.00840$       &  $0.0594$      &  $0.0113$      &  $0.00847$       &  $0.00855$    \\    
MathSciNet                   &  $332,689$        &  $4.93$      &  $0.103$         &  $0.0110$        &  $0.0103$        &  $1.56$        &  $0.0347$      &  $0.0277$        &  $0.0291$     \\    
Social 1                     &  $67$             &  $4.24$      &  $0.103$         &  $0.0486$        &  $0.0418$        &  $0.975$       &  $0.292$       &  $0.179$         &  $0.223$      \\    
Cond-Mat, 1993-2003          &  $21,363$         &  $8.55$      &  $0.125$         &  $0.0103$        &  $0.00947$       &  $7.41$        &  $0.0309$      &  $0.0264$        &  $0.0275$     \\    
Protein 1                    &  $95$             &  $4.48$      &  $0.129$         &  $0.0723$        &  $0.0670$        &  $0.314$       &  $0.384$       &  $0.187$         &  $0.232$      \\    
Cond-Mat, 1995-1999          &  $13,861$         &  $6.44$      &  $0.157$         &  $0.0163$        &  $0.0146$        &  $3.34$        &  $0.0509$      &  $0.0400$        &  $0.0424$     \\    
College football             &  $115$            &  $10.7$      &  $0.162$         &  $0.00977$       &  $0.00967$       &  $1.50$        &  $0.124$       &  $0.0928$        &  $0.102$      \\    
Cond-Mat, 1995-2003          &  $27,519$         &  $8.44$      &  $0.166$         &  $0.00917$       &  $0.00847$       &  $6.09$        &  $0.0293$      &  $0.0248$        &  $0.0258$     \\    
US Patents                   &  $3,764,117$      &  $8.77$      &  $0.168$         &  $0.0103$        &  $0.0100$        &  $8.05$        &  $0.0113$      &  $0.00885$       &  $0.00899$    \\    
Facebook links               &  $63,392$         &  $25.8$      &  $0.177$         &  $0.00143$       &  $0.00140$       &  $25.8$        &  $0.00810$     &  $0.00754$       &  $0.00762$    \\    
Cond-Mat, 1995-2005          &  $36,458$         &  $9.42$      &  $0.177$         &  $0.00814$       &  $0.00761$       &  $12.7$        &  $0.0223$      &  $0.0195$        &  $0.0201$     \\    
Hep-Th, 1995-1999            &  $5,835$          &  $4.74$      &  $0.185$         &  $0.0523$        &  $0.0523$        &  $3.70$        &  $0.0913$      &  $0.0554$        &  $0.0587$     \\    
AstroPhys, 1993-2003         &  $17,903$         &  $22.0$      &  $0.201$         &  $0.00447$       &  $0.00432$       &  $18.9$        &  $0.0117$      &  $0.0106$        &  $0.0108$     \\   
Protein 2                    &  $53$             &  $4.64$      &  $0.209$         &  $0.0536$        &  $0.0500$        &  $0.722$       &  $0.305$       &  $0.172$         &  $0.210$      \\    
Facebook wall                &  $43,953$         &  $8.30$      &  $0.216$         &  $0.00229$       &  $0.00214$       &  $7.86$        &  $0.0277$      &  $0.0252$        &  $0.0261$     \\    
Dublin                       &  $410$            &  $13.5$      &  $0.226$         &  $0.0263$        &  $0.0261$        &  $3.62$        &  $0.0601$      &  $0.0428$        &  $0.0448$     \\    
Actor coll. net.             &  $374,511$        &  $80.2$      &  $0.226$         &  $0.000600$      &  $0.000599$      &  $429$         &  $0.00120$     &  $0.00118$       &  $0.00118$    \\    
Astrophysics                 &  $14,845$         &  $16.1$      &  $0.228$         &  $0.00504$       &  $0.00494$       &  $5.64$        &  $0.0155$      &  $0.0135$        &  $0.0138$     \\    
PGP                          &  $10,680$         &  $4.55$      &  $0.238$         &  $0.0166$        &  $0.0163$        &  $4.25$        &  $0.0301$      &  $0.0236$        &  $0.0243$     \\    
Hep-Th, 1993-2003            &  $8,638$          &  $5.74$      &  $0.239$         &  $0.0312$        &  $0.0312$        &  $8.03$        &  $0.0669$      &  $0.0322$        &  $0.0333$     \\    
Reactome                     &  $5,973$          &  $48.8$      &  $0.241$         &  $0.00414$       &  $0.00413$       &  $27.2$        &  $0.00550$     &  $0.00481$       &  $0.00483$    \\    
Flickr                       &  $105,722$        &  $43.8$      &  $0.247$         &  $0.00105$       &  $0.00105$       &  $101$         &  $0.00170$     &  $0.00162$       &  $0.00163$    \\    
E. Coli, transcription       &  $97$             &  $4.37$      &  $0.412$         &  $0.0854$        &  $0.0807$        &  $0.327$       &  $0.328$       &  $0.153$         &  $0.184$      \\    
Hep-Ph, 1993-2003            &  $11,204$         &  $21.0$      &  $0.630$         &  $0.00389$       &  $0.00389$       &  $153$         &  $0.00450$     &  $0.00408$       &  $0.0041$     \\   
GR-QC, 1993-2003             &  $4,158$          &  $6.46$      &  $0.639$         &  $0.0209$        &  $0.0209$        &  $7.49$        &  $0.0273$      &  $0.0219$        &  $0.0225$     \\    

\hline\hline

  \end{longtable}
\end{center}
\end{widetext}



%

\end{document}